# Spherical Phase Metalens: Intrinsic Suppression of Spherical Aberration via Equiphase Surface Modulation

Xiaohui Yang, Lei Yang*, Xinhui Lu, Yu Guo
School of Mechanical Engineering, Jiangsu University, Zhenjiang 212013, China
* leiyang@ujs.edu.cn

**Abstract** Recent progress in large-scale metasurfaces requires phase profiles beyond traditional hyperbolic designs. We show hyperbolic phase distributions cause spherical aberration from mismatched light propagation geometry and unrealistic phase assumptions. By analyzing metalens fundamentals via isophase surfaces, we develop a spherical phase profile based on spherical wavefront theory. This method prevents spherical aberration, essential for wide-aperture metalens. Simulations prove superior focusing: spherical phase reduces full width at half maximum FWHM by 7.3% and increases peak relative field intensity (PRFI) by 20.4% versus hyperbolic designs at 31.46 micron radius. Spherical phase maintains consistent focusing across radii, while hyperbolic phase shows strong correlation (R squared=0.95) with aberration. We also propose a normal vector tracing metric to measure design aberrations. This work establishes a scalable framework for diffraction-limited metalens.
**Keywords** Metalens, Spherical phase profile, Isophase surface modulation, Spherical aberration elimination, FDTD simulations, Wavefront propagation

This revision incorporates methodological refinements:
1.FDTD parameters: with evanescent wave exclusion justified by >50 dB attenuation.
2.Convergence: 40× mesh coarsening (0.00025→0.01 um) validates <1.2% variation.
3.Code: MATLAB algorithms for aberration quantification now public [GitHub].

## 1.Introduction

Metamaterials are artificially engineered composite materials that exhibit electromagnetic properties not found in natural materials, enabling unprecedented control over electromagnetic wave manipulation. As the two-dimensional counterparts of metamaterials, metasurfaces utilize periodic arrays of subwavelength structural units on planar substrates to generate tailored electromagnetic responses for diverse device functionalities. Metasurfaces enable unprecedented multidimensional wavefront control, spanning wavelength [1], polarization [2], and angular [3] degrees of freedom, while maintaining ultracompact footprints. The operational principle of metasurfaces relies on the phase modulation imparted by structural units at the interface, which confers advantages such as reduced thickness, lower optical losses [4], and higher integration density [5] compared to conventional optical devices. Representative applications include high-numerical-aperture achromatic metalens [6,7] and polarization-multiplexed metasurface holograms [2,8]. High-efficiency metasurfaces predominantly employ high-index dielectric or semiconductor materials—such as titanium dioxide [9,10], gallium nitride [11,12], and silicon [13,14]. Silicon-based metasurfaces further benefit from compatibility with modern complementary metal-oxide-semiconductor (CMOS) fabrication processes, positioning the technology for scalable manufacturing.

The design of metasurface devices primarily relies on phase gradient modulation. This concept originates from the generalized laws of refraction and reflection proposed by Yu et al. [15] in 2011, which were experimentally validated through beam deflection using V-shaped structural unit arrays. Similarly, metalens designs depend on phase gradient modulation. The foundational work on phase distribution in metalens can be traced to the hyperbolic phase profile introduced by Capasso's group at Harvard University in 2016 [16], hyperbolic phase profile ($\varphi = -\frac{2\pi}{\lambda}\left(\sqrt{r^2 + f^2} - f\right)$). Subsequent studies have advanced the control of focal position and field of view in metalens.

Research on focal spatial positioning encompasses two main categories: achromatic metalens and multifocal metalens. Achromatic metalens require simultaneous compensation of both phase dispersion and group delay across different angular frequencies to ensure temporal synchronization of all spectral components at the focal point. In 2018, Wang et al. [17] demonstrated a gallium nitride-based transmission-phase metalens with a numerical aperture (NA) of 0.106 operating across 400–660 nm. In the same year, Chen et al. [18] achieved a resonant-phase titanium dioxide metalens with NA = 0.2 over 470–670 nm. These works spurred investigations into both high-aspect-ratio simple units and multi-component complex units. Recently, Chen et al. [19] proposed a quasi-achromatic metalens in 2023, extending the achromatic bandwidth by introducing a $2\pi/\Delta\omega$-periodic group delay modulation. Their simulations achieved a 1 mm-radius metalens with NA = 0.55 spanning 400–1500 nm. For multifocal metalens, discrete multi-channel configurations are common. For instance, in 2022, Yang's group at Jiangsu University [20] developed a triple-focal, multi-channel metalens based on phase-change materials and circular polarization to analyze cascaded focal positions in a system of two single-layer hyperbolic metalens. The field of view in metalens is predominantly limited by coma aberration. Existing research generally establishes that hyperbolic phase profiles inherently eliminate spherical aberration and distortion [21]. Leveraging this property, efforts to optimize the field of view focus on balancing coma aberration through precise phase distribution design. The current strategy for aberration correction introduces an effective aperture dependent on the incident angle. This design ensures that light with different incident angles generates phase mutations in distinct regions of the metalens. In 2017, Groever et al. [22] introduced polynomial terms for optimization and employed ray tracing to correct the hyperbolic phase profile, achieving a double-layer metalens with a numerical aperture (NA) of 0.44 and a field of view of 50°. In the same year, Pu et al. [23] proposed a quadratic phase profile, quadratic phase profile ($\varphi = -\frac{\pi}{\lambda} \cdot \frac{r^2}{f}$), when combined with the phase of obliquely incident plane waves, achieved an adaptive optical axis that shifts with the incident angle and a virtual aperture effect. In 2021, Lassalle et al. [24] realized a secondary-phase metalens with a diameter of 500 um and a field of view of 120°.

In addition to phase research, the theory of exploring the upper limit of performance of metamaterial lenses is also constantly evolving [25–29]. With the development of deep learning technology, the reverse design method of metamaterial lenses plays an important role in areas where forward design is difficult to solve [3,30–35]. However, regardless of subsequent innovations by researchers, all designs remain fundamentally rooted in the hyperbolic phase framework. This foundational dependency has resulted in inherent limitations across all metalens phase profiles developed since the introduction of hyperbolic phase principles.

In studying metalens phase distributions, researchers have inherited assumptions from traditional optics: it is postulated that light vertically incident on metalens structural units undergoes directional changes upon emission. Contrastingly, simulated electric field distributions of the structural units confirm that the propagation direction of emitted light remains aligned with the incident direction; therefore, directional modulation occurs exclusively at the phase gradient interface (i.e., the spherical equiphase surface in air). This discrepancy introduces systematic phase errors, particularly spherical aberration, in existing metalens designs. Such aberrations become especially pronounced in single-layer, centimeter-scale hyperbolic-phase metalens, where spherical aberration cannot be neglected.

Building on the premise that metalens nanostructures preserve the propagation direction of transmitted light, we re-examine the imaging mechanism of metalens. Our analysis reveals that light at any radial position on the metalens propagates normal to its surface until intersecting the equiphase surface of the outgoing wavefront. This capability enables direct modulation of the optical field by engineering the equiphase surface of transmitted light rather than the physical lens structure Kirchhoff diffraction theory further demonstrates that, under idealized conditions—where the equiphase surface forms a perfect sphere and the system is free from imperfections—the focal point is theoretically devoid of geometric aberrations such as spherical aberration, coma, and astigmatism. Crucially, this equiphase surface modulation paradigm diverges fundamentally from conventional optical systems, where such direct wavefront control is inherently unattainable.

This study investigates metalens through the analysis of equiphase surfaces and defines the generalized phase profile of a metalens as the phase difference between the outgoing spherical equiphase surface and the incident planar equiphase surface. We further propose a spherical phase formula derived from this definition. To evaluate imaging quality, we propose a normal-tracing method based on equiphase surface propagation and provide a quantitative formula for characterizing spherical aberration in metalens under this framework. This method enables performance assessment during the initial design phase. Validation is achieved via three finite-difference time-domain (FDTD) simulation studies: (1) fixed radius with variable numerical aperture ($NA$), (2) fixed focal length with variable radius($r$), and (3) fixed NA with variable radius($r$). All simulation results demonstrate that the spherical phase profile achieves narrower full width at half maximum (FWHM) and higher peak relative field intensity (PRFI) compared to hyperbolic counterparts. Compared to conventional hyperbolic phase profiles, the spherical phase profile proposed in this work offers superior suitability for designing and fabricating centimeter-scale metalens. By focusing on theoretical innovations in phase distribution, this framework integrates seamlessly with existing metalens architectures while introducing novel design paradigms to advance future research.

## 2. Principles and Methods
### 2.1 Limitations of Hyperbolic Phase Profiles

All incident light discussed in this chapter comprises collimated beams with vertical incidence, characterized by planar equiphase surfaces. Metasurface devices function as phase modulation elements, where metamaterial lenses manipulate light via phase mutations. For modulated beams, the alteration in propagation direction occurs precisely at the equiphase surface within free space, defined by these phase mutations. This principle is exemplified in the seminal work of Yu et al. [15] (2011), as illustrated in Fig. 1. Specifically, Fig.1(b) designates $S_{in}$ and $S_{out}$ as the equiphase surfaces of the incident and transmitted metasurfaces, respectively, with $S_{in\_1}$ and $S_{out\_1}$ representing the surfaces closest to the metasructure.

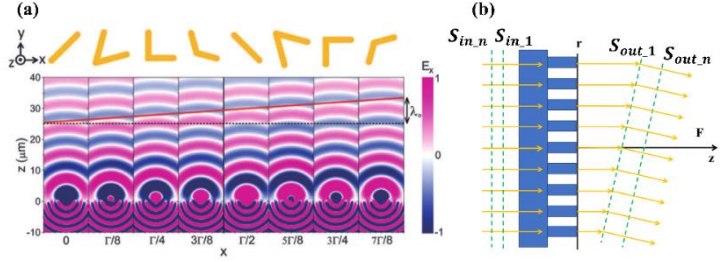

**Fig.1.** Metasurface-based beam deflection device. (a) Schematic of the phase-modulated beam deflection mechanism adapted from Yu *et al*. [15]. (b) Simplified representation of the incident and outgoing equiphase surfaces.

The imaging mechanism of single-layer metalens originates from the hyperbolic phase profile first introduced by Capasso et al. [16] in 2016, which they described as enabling the device "to function like a spherical lens." The mathematical expression of this hyperbolic phase is provided in Eq. (1), while its schematic implementation is illustrated in Fig. 2.

$$\varphi_{nf}(x,y) = -\frac{2\pi}{\lambda_d}\left(\sqrt{x^2 + y^2 + f^2} - f\right) \tag{1}$$

Here, $\lambda_d$ represents the operating wavelength of the metamaterial lens, and $\varphi_{nf}$ denotes the phase mutation generated by the lens. Fig. 1(a) depicts the optical path diagram of the metamaterial lens, where the green curve indicates the equiphase surface in free space, and yellow solid lines represent the light rays. As observed in Fig. 1(a), the propagation direction of light changes before reaching the equiphase surface (at the interface between the metamaterial lens and free space) without undergoing phase mutation modulation. This indicates that the phase mutation exerts no actual effect, thereby contradicting the fundamental beam deflection mechanism of metasurfaces. This implies that the phase mutation contributes negligibly to the beam deflection, which directly contradicts the foundational principle of metasurfaces. Fig. 2(b) and 2(c) illustrate the structural unit schematics of the hyperbolic-phase metamaterial lens and the metasurface beam deflection device, respectively.

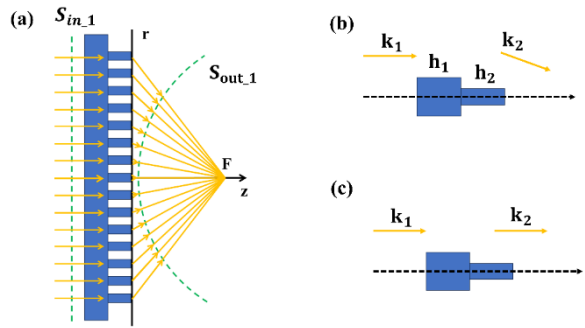

Fig. 2. illustrates the hyperbolic phase profile, where k₁ and k₂ denote the propagation directions of light in air. (a), (b), and (c) depict the hyperbolic phase distribution, the theoretical output direction of light from the structural unit under this phase profile, and the actual output direction of light from the structural unit, respectively.

The optical path of the hyperbolic phase profile in Fig. 2(a) closely resembles that of a traditional plano-convex lens collimating parallel light (Fig. 3), suggesting a conceptual analogy between the metasurface phase design and conventional refractive optics.

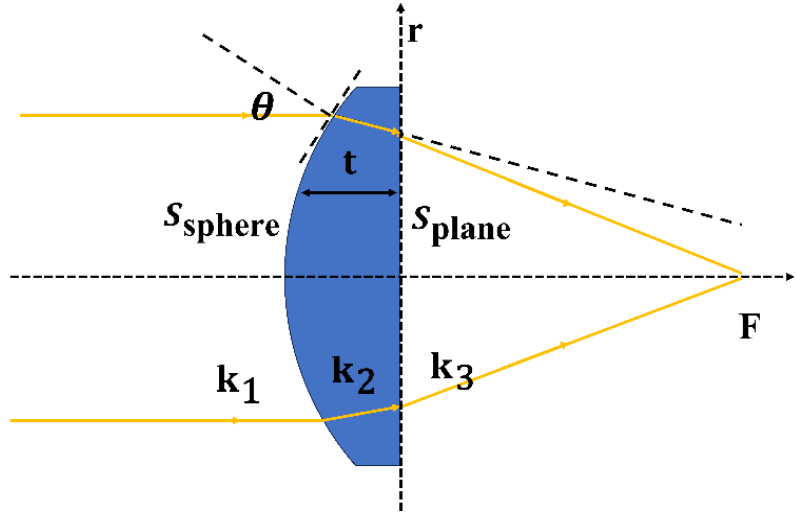

Fig. 3. Optical path diagram of a plano-convex lens, where $k_1$、$k_2$, and $k_3$ denote the light propagation directions within three distinct uniform media.

The optical path in Fig. 2(a) directly mirrors that of the plano-convex lens configuration in Fig. 3, depicting light exiting the lens' planar surface into air. However, this comparison overlooks two critical distinctions: First, plano-convex lenses operate at a macroscopic scale—even thin lenses exhibit thicknesses (t) on the millimeter to centimeter scale, far exceeding optical wavelengths. In contrast, metasurface lenses function at a microscopic scale, where both the substrate thickness ($h_1$) and structural unit height ($h_2$) remain uniform across the lens surface (r). Second, light propagating through the plano-convex lens' planar interface (Fig. 3) possesses a non-zero incidence angle, enabling secondary modulation of its direction. Conversely, the metasurface in Fig. 2(a) employs normal incidence (vertically aligned illumination), eliminating both the curvature-dependent thickness variations (t) and incidence angles ($\theta$) inherent to conventional lenses. Consequently, metasurface structural units cannot modulate light via macroscopic geometric features (e.g., spherical surfaces) or planar interfaces with tilted propagation, fundamentally distinguishing their mechanism from traditional refractive optics.

Therefore, we conclude that the propagation direction of light remains unaltered upon passing through the structural units of the metasurface lens; these units solely introduce a localized phase delay. The resultant beam deflection is governed exclusively by the equiphase surface formed via the phase gradient in free space, rather than direct modulation by the structural units. The actual propagation directions of light before and after interacting with the structural units are explicitly depicted in Fig. 2(c).

Therefore, the hyperbolic phase profile exhibits a systematic phase error originating from the misconceptions depicted in Fig. 2(b), where the structural units were erroneously assumed to directly modulate the propagation direction. Furthermore, this study demonstrates that such systematic phase errors are additionally linked to phase folding effects.

Phase folding, a critical step for matching the phase distribution of a metasurface lens to its structural units' phase response, involves truncating phase values exceeding 2*pi (induced by the lens' physical radius) into the 0~2*pi range. However, conventional approaches neglect the inherent dependence of phase folding on light propagation direction. Physically, this process modifies the effective wavenumber of light while maintaining its spatial position and propagation direction—a prerequisite for wavefront continuity. As demonstrated in Fig. 4, points A, B, and C on the metasurface interface correspond spatially to points D and E on the free-space equiphase surface (shared coordinate r), illustrating how phase folding distorts the hyperbolic phase profile. This distortion originates from a directional mismatch between the light propagation path and the orientation of phase folding. Specifically, the phase folding process fails to account for the alignment of the folded phase gradient with the actual beam trajectory, introducing systematic phase errors.

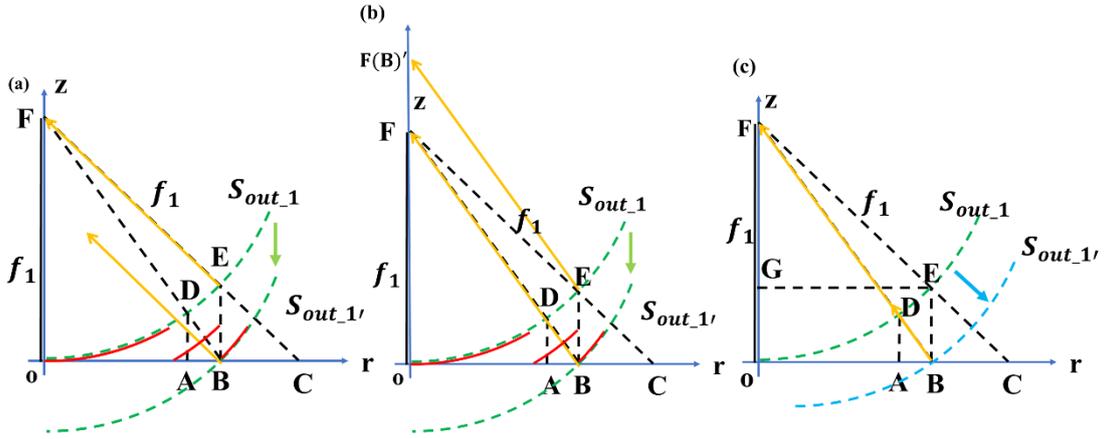

Fig.4. The Effect of Phase Folding on Metalens. (a) Phase Folding Diagram. (b) Hyperbolic Phase Actual Optical Path Diagram. (c) Ideal optical path under the assumption that the hyperbolic phase beam direction aligns with the phase folding orientation.

Fig. 4(a) assumes beam deflection occurs at the equiphase surface, while Fig. 4(b) and 4(c) postulate deflection at the metasurface interface. The red dashed lines depict the folded phase profile after truncation. In the actual phase folding direction (green arrows), the equiphase surface $S_{out_{1'}}$ is derived by translating $S_{out\_1}$ along the negative z-axis, preserving the propagation direction, curvature, and surface normal at spatial position r, as governed by $\varphi_{out_{1'}} = \varphi_{out_1} + 2n*pi$, n ∈ Z). Conversely, the "ideal" folding direction (blue arrows) assumes $S_{out_{1'}}$ and $S_{out\_1}$ are concentric spherical surfaces centered at point F, altering both curvature and surface normals at r. The yellow lines represent surface normals before and after folding: In Fig. 4(a), the surface normal directions at points B and E are identical. In Fig. 4(b), the surface normal directions at points E and B remain the same. In Fig. 4(c), the surface normal directions at points B and D coincide.

Fig. 4(a) illustrates an idealized but flawed interpretation of hyperbolic-phase metasurface lenses, serving as the root of the misconception that such lenses can perfectly collimate plane waves. This error arises from two critical omissions: (1) the failure to recognize that the hyperbolic phase inherently induces beam deflection at the metasurface interface (Fig. 4(b)), and (2) the blind application of the hyperbolic phase formula followed by phase folding, which erroneously assumes beam deflection occurs only at the equiphase surface. By neglecting both the interfacial deflection mechanism and the physical constraints of phase folding (which requires preserving spatial position and propagation direction during wavenumber modulation), researchers incorrectly conclude that hyperbolic-phase metasurfaces achieve aberration-free imaging.

The misconception in Fig. 4(a) stems from a flawed premise: the hyperbolic phase formula is applied directly, followed by phase folding, without recognizing that the hyperbolic phase inherently induces beam deflection at the metasurface interface (Fig. 4(b)). Critically, phase folding requires the assumption that beam deflection occurs only at the equiphase surface—a physical constraint demanding preservation of both spatial position and propagation direction during wavenumber modulation. By simultaneously ignoring the interfacial deflection and violating this prerequisite, researchers erroneously conclude that hyperbolic-phase metasurfaces achieve perfect plane-wave collimation.

Fig. 4(b) reveals the actual optical path under phase folding. At the metasurface interface (point B), the hyperbolic phase alters the propagation direction, defining the surface normal as $\overrightarrow{L_{BF}}$. Since phase folding preserves this normal direction, light emitted from point E converges not at F, but at $F'(B)$ along $\overrightarrow{L_{BF}}$. The propagation distance $|\overrightarrow{L_{BF}}|$ exceeds the focal length $|\overrightarrow{L_{DF}}| = f_1$, violating the ideal imaging condition ($|\overrightarrow{L_{BF}}| > f_1$) and introducing spherical aberration. This quantifies the geometric inconsistency between the designed phase profile and physical light propagation.

Fig. 4(c) proposes a remedial strategy: When folding the phase distribution while explicitly accounting for interfacial beam deflection, the equiphase surface $S_{out_{1'}}$ is forced to share the same spherical center as $S_{out\_1}$. This ensures the propagation directions at points B and D align, enabling focused convergence at F. However, this geometric constraint breaks the conventional phase relation $\varphi_{out_{1'}} = \varphi_{out_1} + 2n*pi$, $n \in Z$), necessitating a specialized derivation of the $S_{out_{1'}}$ phase profile to reconcile wavefront continuity with the modified beam trajectory.

Unlike the spherical phase profile, which precisely modulates the propagation direction of emitted light to achieve diffraction-limited focusing, the hyperbolic phase profile fails to adapt its phase distribution to the geometric requirements of focal convergence. This intrinsic mismatch between the hyperbolic phase and light propagation geometry results in spherical aberration, even at low numerical apertures (NA), as the phase profile cannot correct for deviations in the beam trajectory.

**2.2 Derivation of the Spherical Phase Profile**

As established earlier, the nanostructures of a metalens do not alter the propagation direction of transmitted light—a phenomenon inconceivable in conventional optics. Specifically, collimated light incident perpendicularly to the metalens surface propagates uniformly along the normal direction across all radial positions (r) until reaching the equiphase surface of the outgoing light. For ideal focusing, this equiphase surface must conform to a spherical geometry. This unique property enables precise modulation of the outgoing spherical wavefront through metasurface engineering. Under the Kirchhoff diffraction framework, if the equiphase surface is strictly spherical and the system is free from physical imperfections, the resulting focal point theoretically exhibits zero geometric aberrations (e.g., spherical aberration, coma, and astigmatism). Fig. 5 schematically illustrates the equiphase surfaces for planar, divergent spherical, and convergent spherical wavefronts.

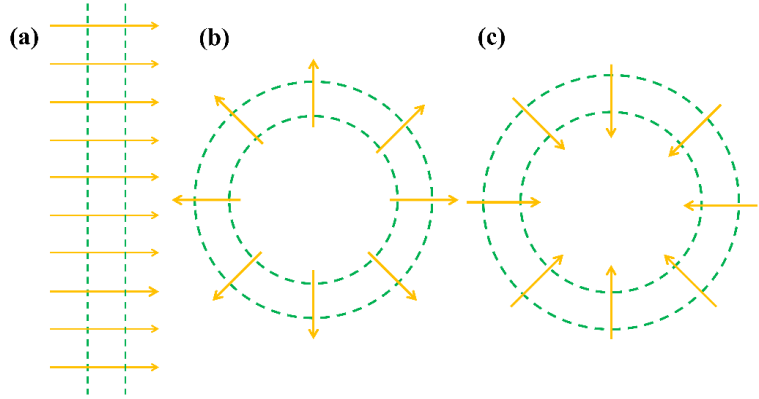

Fig. 5. Schematic diagrams of equiphase surfaces. (a) Planar equiphase surface. (b) Divergent spherical equiphase surface. (c) Convergent spherical equiphase surface.

An equiphase surface (or wavefront) is defined as a surface composed of spatial points sharing identical phase. At any point on this surface, the wavevector k is perpendicular to the equiphase surface and aligns with the wave propagation direction. According to Huygens' principle, the propagation direction of light in a homogeneous medium always follows the surface normal of the equiphase surface. For a convergent equiphase surface in air, the focal position can thus be determined by tracing the normals across the surface.

The directional modulation of transmitted light in a metalens is inherently localized to the spherical equiphase surface $S_{out\_1}$ (the output spherical wavefront), as schematically detailed in Fig. 6(a).

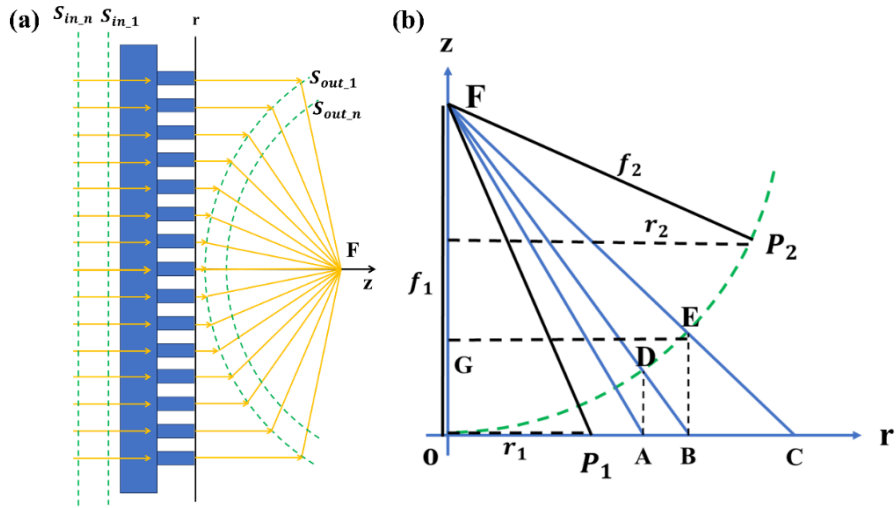

Fig. 6. Schematic ray paths of spherical-phase metalens. (a) Proposed spherical phase profile design for the metalens. (b) Comparative analysis of hyperbolic and spherical phase profiles under identical focal lengths ($f_1 = f_2$). Points A, B, and C follow the hyperbolic phase distribution, while points D and E adhere to the converging spherical phase profile. Crucially, points A and D (as well as B and E) share identical radial positions (r), enabling direct comparison of their phase-dependent focusing behaviors.

Fig. 6(a) demonstrates the light propagation mechanism of the metalens governed by the proposed spherical phase profile. Equiphase surfaces are depicted as green dashed lines, with yellow solid arrows indicating light propagation directions. Here, $S_{in}$ and $S_{out}$ represent the incident planar and outgoing spherical equiphase surfaces, respectively. Specifically,

$S_{in\_1}$ denotes the planar equiphase surface directly adjacent to the metalens, while $S_{out\_1}$ corresponds to the nearest spherical equiphase surface generated by the nanostructures. As both $S_{in}$ and $S_{out}$ exist in a homogeneous air medium, light propagates rectilinearly between successive equiphase surfaces (e.g., from $S_{in\_1}$ to $S_{in}$ and $S_{out\_1}$ to $S_{out}$).

The imaging process of the metalens proceeds as follows: Collimated light from the planar equiphase surface $S_{in\_1}$ is normally incident on the metalens. Each nanostructure functions as a secondary Huygens source, producing transmitted light that collectively forms a new wavefront at $S_{out\_1}$ (in air). The compensated phase introduced by the metalens corresponds to the phase difference between $S_{in\_1}$ and $S_{out\_1}$. Consequently, the generalized phase profile of the metalens is expressed as:

$$\varphi(r) = S_{out_1}(r) - S_{in_1}(r) \tag{2}$$

Here, $S_{in_1}(r)$ and $S_{out_1}(r)$ represent the equiphase surfaces of the incident and transmitted light, respectively. When $S_{out_1}(r)$ adopts a spherical geometry, the spherical phase profile of the metalens is governed by Eq. (3). A detailed derivation of this phase formula is provided in Supplementary Section 1, Fig. S2.

$$\varphi_{sphere}(r) = -\frac{2\pi}{\lambda_d}\left(f - \sqrt{f^2 - r^2}\right) \tag{3}$$

Fig. 6(b) provides a schematic comparison of the hyperbolic and spherical phase profiles, using point B as an example. For the hyperbolic phase profile, the optical path is represented by the vector $\overrightarrow{L_{BD}}$, while the spherical phase profile corresponds to the vector $\overrightarrow{L_{BE}}$. These vectors differ in both magnitude and direction, highlighting the fundamental differences in wavefront modulation between the two phase profiles.

**2.3 Characterization of Spherical Aberration in Metalens**

For an ideal converging spherical equiphase surface, all points on the surface propagate along their respective normals to converge at the focal point, inherently eliminating spherical aberration. Thus, a metasurface lens exhibiting a spherical output equiphase surface achieves theoretically aberration-free focusing. This principle enables phase quality evaluation through normal vector tracing: the spherical aberration at each point of the equiphase surface quantitatively reflects the metasurface's phase fidelity. In this work, we propose, for the first time, a normal vector tracing method to assess metasurface lens performance, accompanied by a dedicated spherical aberration metric (Eq. (8)). Crucially, this approach allows pre-fabrication evaluation of imaging quality based solely on the phase profile—prior to numerical simulations or experimental validation—providing a predictive tool for optimizing metasurface designs.

Spherical aberration, caused by light rays from different radial positions (r) on the lens converging to distinct focal points, is inherently spatially dependent. This aberration can be quantified using the Focal Length Absolute Error (FLAE)—the deviation between the actual and designed focal lengths at each point on the outgoing equiphase surface $S_{out_1}$ To derive the FLAE metric, we first express the phase profiles of Eq. (1) and Eq. (3) as equivalent optical path difference (OPD) terms, given by Eq. (4) and Eq. (5), respectively.

$$L_{hyperbolic} = \sqrt{r^2 + f^2} - f \tag{4}$$

$$L_{sphere} = f - \sqrt{f^2 - r^2} \tag{5}$$

Next, surface normals are determined for each point on the optical path difference (OPD) profiles defined by Eq. (4) and Eq. (5). Extending these normals to intersect the optical axis yields position-dependent focal points $F'(r)$ Using the designed focal length F of the metalens, the Focal Length Absolute Error (FLAE) is defined as:

$$FLAE(r) = abs(F'(r) - F) \tag{6}$$

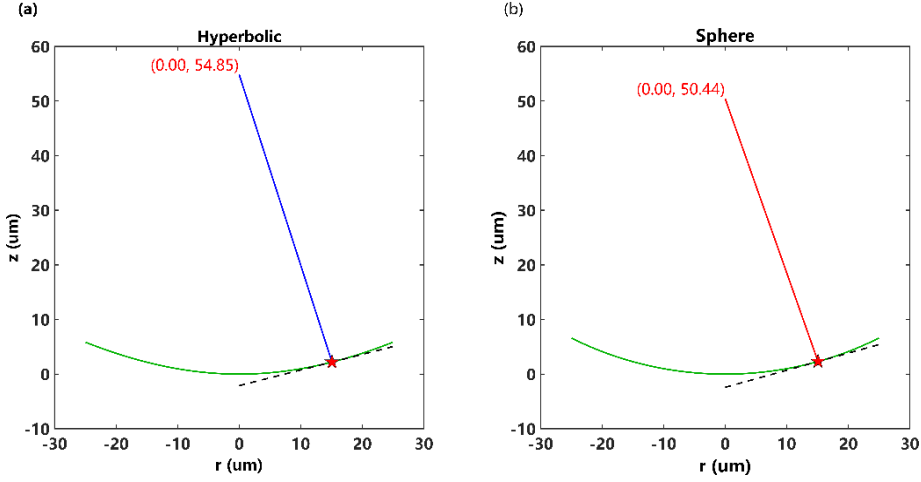

Fig. 7. Normal-tracing diagrams for hyperbolic and spherical phase profiles. The star symbol (★) marks a randomly selected radial position (r) on the metalens. (a) Hyperbolic phase profile: The focal length at ★ is 54.85 um. (b) Spherical phase profile: The focal length at ★ is 50.44 um.

Fig.7 and 8 characterize a metalens with a radius R=25.22 um and a designed focal length F=50.44 um, operating at a wavelength of 1.1 um. The unit cell period is 0.52μm, with the marker (★) indicating a representative spatial position. Fig. 7(a) and 7(b) compare the focal lengths derived from ray-tracing simulations at this position for hyperbolic and spherical phase profiles, respectively. The hyperbolic phase profile yields a focal length of 54.85 um, deviating from the designed value by 8.7 %. In contrast, the spherical phase profile produces a focal length of 50.44 um, closely matching the intended design value. The results in Fig. 7 (a) are consistent with the analysis in Fig. 4 (b).

The diffraction limit (DL) of a metalens, which governs its minimum resolvable feature size, is determined by its inherent physical properties: the operating wavelength λ, radius R, and focal length F. For a diffraction-limited metalens, this relationship is expressed as:

$$DL = \frac{1.22 * \lambda}{NA} = \frac{1.22 * \lambda}{\sin\left(\tan^{-1}\frac{R}{F}\right)} \tag{7}$$

Given the subwavelength operational regime of metalens, we introduce the focal length absolute error (FLAE(r)) to visualize deviations and systematically assess spherical aberrations across variations in lens radius (R), focal length (F), and wavelength (λ). To unify this analysis, FLAE(r) is normalized to the intrinsic diffraction limit (DL) of the metalens. The spherical aberration is then quantified as a dimensionless metric:

$$\text{Spherical Aberration}(r) = \frac{FLAE(r)}{DL} = \frac{abs(F'(r) - F)}{DL} \tag{8}$$

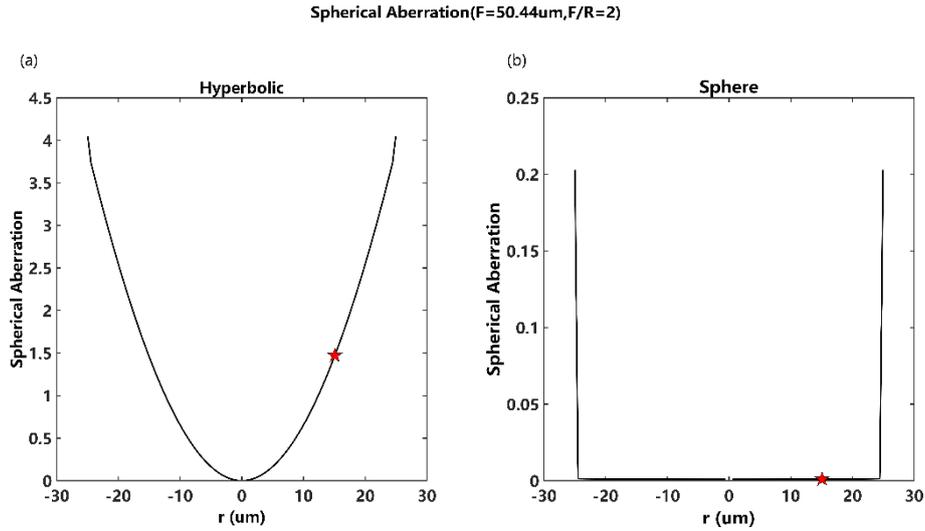

Fig. 8. Spherical aberration in the metalens. (a) Hyperbolic phase-induced spherical aberration. (b) Spherical phase-induced spherical aberration.

Fig. 8(a) and 8(b) showcase the spatial distribution of spherical aberration across the metalens. For the hyperbolic phase profile, spherical aberration increases monotonically with radial position r. In contrast, the spherical phase profile maintains zero aberration across the entire lens aperture, with finite values observed only at the boundary due to edge effects. These results align with the theoretical predictions of this study, confirming that the proposed spherical phase profile eliminates the spherical aberration inherent to hyperbolic designs. This advancement enables the design of large-scale metalens with diffraction-limited performance. Furthermore, the fixed focal length in Fig. 7 and the direct dependence of numerical aperture (NA) on lens radius demonstrate that the spherical phase profile is equally effective for high-NA metalens.

By conducting normal vector-based ray-tracing simulations for both hyperbolic phase profiles and spherical phase profiles and analyzing the spherical aberration of the metalens via MATLAB, two key conclusions can be drawn:

Conclusion 1: The hyperbolic phase profile introduces spherical aberration, while the spherical phase profile remains aberration-free. Consequently, the spherical phase profile is more suitable for designing and studying large-scale metalens (e.g., centimeter-scale).

Conclusion 2: Geometric derivations of the spherical phase profile reveal a fundamental constraint between focal length (f) and radius (r): the focal length must exceed the radius (f>r). This relationship imposes a theoretical upper limit on the numerical aperture (NA) of aberration-free metalens (NA = 0.707). Thus, single-layer metalens with NA values surpassing 0.707 will inherently exhibit spherical aberration.

## 3. Simulations and Discussion

Simulations were conducted using FDTD Solutions 2023 R2 and MATLAB R2020b.

The metalens simulation employed a near-field to far-field diffraction approach with:
- Minimum mesh step: 0.00025 um ($\lambda$/154 in silicon)
- Convergence: Verified through 40× coarsening (3.3 Section for critical case analysis)
- Far-field: farfielddexact3d function with 10 points/um sampling
- Evanescent waves: Excluded based on >50 dB attenuation in silicon (Supplementary

Section 7, Table S2)

Relative field intensity (RFI) is defined as |E|² in arbitrary units.

To demonstrate the superiority of the spherical phase profile over the hyperbolic counterpart, all simulations utilized conventional low-aspect-ratio brick-shaped silicon nanostructures on a silicon dioxide substrate. Design parameters included a period P = 0.52 µm, unit height h = 0.75 µm, operating wavelength $\lambda$ = 1.1 µm, and x-polarized incidence. A representative nanostructure with dimensions Lx = 0.3 µm and Ly = 0.1 µm is illustrated in Fig. 9, while Fig. S3 (Supplementary Section 2) details the electric field propagation directions of incident and transmitted light. Transmittance, and phase libraries of the nanostructures are provided in Fig. S4 (Supplementary Section 2).

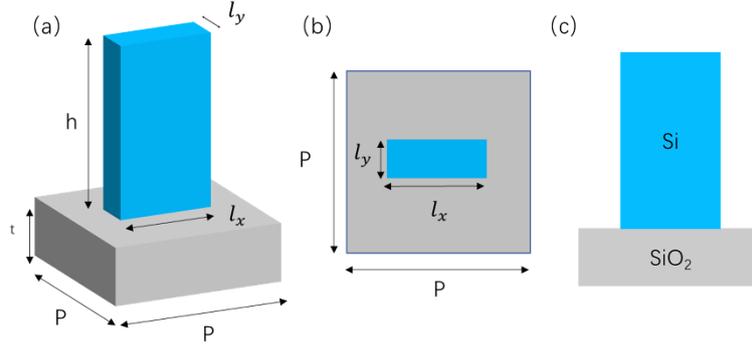

Fig. 9. Schematic diagrams of the nanostructure unit. (a) 3D view. (b) top view. (c) front view.

To validate the inherent spherical aberration in the hyperbolic phase profile and confirm the theoretical absence of such aberration in the spherical phase profile, we conducted three comparative simulations: (1) investigating the focusing performance of the metalens under varying numerical aperture (NA) with a fixed radius R; (2) investigating the effect of different radii R on focusing efficiency at a fixed focal length F; and (3) investigating the influence of varying radii R on focal quality under a fixed NA.

In these simulations, focal length (F), numerical aperture (NA), and radius (R) were systematically varied. To standardize the comparison of focusing performance between hyperbolic and spherical phase profiles, we introduced the relative spot size (RSS), defined as:

$$\text{RSS} = \frac{\text{FWHM}_{\text{Simulation}}}{\text{DL}_{\text{design}}} \tag{9}$$

Similarly, the focal length relative error (FLRE) is expressed as:

$$\text{FLRE} = \frac{\text{FLAE}}{F} = \frac{\text{abs}(f_{\text{Simulation}} - F_{\text{design}})}{F_{\text{design}}} \tag{10}$$

### 3.1 Investigating the Focusing Performance of Metalens at Fixed Radius with Variable Numerical Aperture *NA*

Fig. 10 compares the focusing performance of spherical and hyperbolic phase profiles under varying numerical apertures (NA, determined by focal length F), with corresponding electric field distributions at the focal plane provided in Supplementary Section 3, Fig. S5. In these simulations, the metalens retains a fixed radius R = 25.22 µm, a 97×97 array configuration, and a unit period of 0.52 µm. The operating wavelength is $\lambda$ = 1.1 µm, and the focal length F is adjusted systematically relative to R (F/R = 2–9) to evaluate the impact of varying *NA* on focusing efficiency.

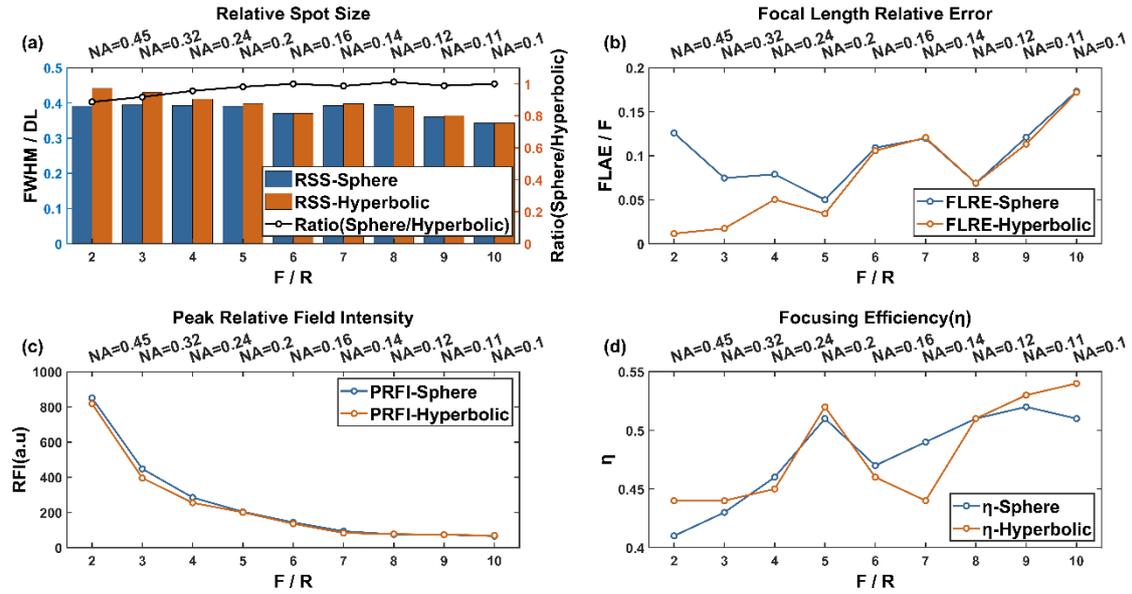

Fig.10. Comparative analysis of spherical and hyperbolic phase profiles under varying focal lengths (numerical apertures). (a) Relative spot size (RSS) versus numerical aperture (*NA*) for both phase designs. (b) Focal length relative error (FLRE) versus *NA*. (c) Peak relative field intensity (PRFI) at the focal point versus *NA*. (d) Focusing efficiency (η) versus *NA*.

Fig. 10 employs dual horizontal axes: the lower axis represents focal length F, while the upper axis shows the corresponding numerical aperture (*NA*). Fig. 10 (a) features dual vertical axes: the left axis plots $RSS_{sphere}$ and $RSS_{hyperbolic}$, while the right axis displays their ratio $RSS_{sphere}/RSS_{hyperbolic}$. Key observations include:

1. At identical NA (or F), the spherical phase profile exhibits a lower relative spot size (RSS) and higher peak relative field intensity (PRFI) compared to the hyperbolic profile ($RSS_{sphere} < RSS_{hyperbolic}$, $PEI_{sphere} > PEI_{hyperbolic}$; see Fig. 10 (a,c)).

2. The spherical and hyperbolic phase profiles exhibit closely matched focusing efficiencies (η) across all tested numerical aperture (*NA*) conditions, as shown in Fig. 10d.

3. The $RSS_{sphere}/RSS_{hyperbolic}$ ratio (right axis, Fig. 10a) decreases with increasing NA (or decreasing F), indicating enhanced focusing capability of the spherical phase profile at higher NA values relative to its hyperbolic counterpart.

While Fig. 10 (b) initially appears to indicate poorer performance of the spherical phase profile, our analysis identifies two critical findings:

Focal length relative error (FLRE) universality: Both phase profiles exhibit FLRE (Fig. 10b), demonstrating systematic discrepancies between simulated and designed focal lengths. These discrepancies originate from the fundamental difference in phase implementation: metalens produce discrete phase distributions, in contrast to the continuous phase profiles of conventional glass lenses. During simulations, mismatches between the discrete phase distribution and the phase library introduce phase discretization errors in simulations

Phase-dependent FLRE trend: The spherical phase profile exhibits consistently higher focal length relative error than the hyperbolic counterpart across all NA values ($FLRE_{sphere} > FLRE_{hyperbolic}$). This behavior correlates with normal vector ray-tracing results of equiphase surfaces in Fig. 7: spherical phase profiles produce normal-traced focal positions matching the design focal length (*F*), whereas hyperbolic profiles generate foci deviating from F, forming defocused spots. The apparent contradiction—reduced FLRE in hyperbolic profiles despite their focal spot degradation—likely originates from unoptimized spatial offsets between equiphase surfaces and the metalens plane. As shown in Fig. 6a, the spherical phase calculation assumes zero spacing between the equiphase surface and lens plane, an idealized assumption that inadequately accounts for discrete phase sampling effects in physical implementations.

In summary, fixed-radius (R) comparative simulations reveal the superior beam convergence of spherical phase profiles over hyperbolic counterparts under variable numerical aperture (*NA*) or focal length (*F*) conditions. The spherical profile demonstrates two key advantages: (1) smaller full-width at half-maximum (FWHM), indicating tighter focal spot confinement; and (2) higher peak relative field intensity (PRFI), reflecting enhanced focal field strength. These metrics collectively validate the improved light confinement capability of spherical phase profiles, which stems from their inherent spherical aberration immunity. This performance advantage becomes increasingly pronounced at higher NA values, as evidenced by the declining $RSS_{sphere}/RSS_{hyperbolic}$ ratio with increasing numerical aperture in Fig. 9a.

## 3.2 Investigating the Focusing Performance of Metalens at Fixed Focal Length with Variable Radii *R*

Fig. 11 compares the radius-dependent focusing performance of spherical and hyperbolic phase profiles, with corresponding focal plane electric field distributions provided in Supplementary Section 4, Fig. S6. These simulations employ a fixed focal length F = 50 μm, unit period of 0.52 μm, and operating wavelength λ = 1.1 μm. The radius *R* is varied systematically relative to F (F/R=1.8,2.1,2.4,2.7,3.0) to evaluate metalens performance under different geometric scales.

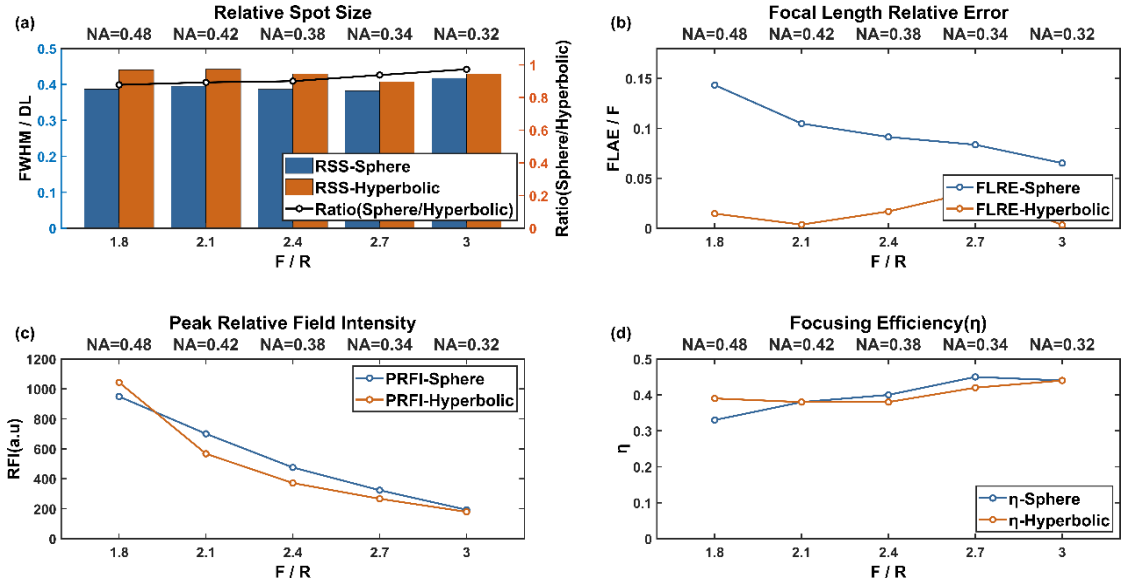

Fig.11. Comparative analysis of spherical and hyperbolic phase profiles under a fixed focal length (F) with varying radius (*R*). (a) Relative spot size (RSS) versus *R*. (b) Focal length relative error (FLRE) versus *R*. (c) Peak relative field intensity (PRFI) versus *R*. (d) Focusing efficiency (η) versus *R*.

The results in Fig. 11 corroborate the trends observed in Fig. 10, with the spherical phase profile consistently demonstrating stronger beam convergence compared to the hyperbolic counterpart. Simulations in Section 3.1 establish the spherical phase's superiority at high numerical apertures, while Section 3.2 validates its scalability advantages for large-scale metalens Given the inherent coupling between key parameters (*NA*, *F*, and *R*), we conducted additional simulations with fixed NA while proportionally scaling both focal length *F* and radius *R*. This parameter decoupling strategy enables systematic evaluation of dimensional scaling effects on spherical-phase metalens focusing performance.

## 3.3 Exploring the Impact of Lens Radius *R* on Focusing Performance at a Constant Numerical Aperture

As shown in Fig. 12, under a fixed numerical aperture (NA = 0.317) with proportionally scaled lens radius, spherical-phase metalens demonstrate consistent superiority in focusing performance over hyperbolic-phase counterparts. Corresponding focal plane electric field distributions are provided in Supplementary Section 5, Fig. S7. All compared metalens maintain identical structural parameters: 0.52 µm unit period, operational wavelength λ = 1.1 µm, and fixed focal-length-to-radius ratio (F/R= 3.0). This configuration isolates geometric scaling effects while preserving wavefront matching conditions, building on the systematic parameter decoupling methodology established in Sections 3.1-3.2.

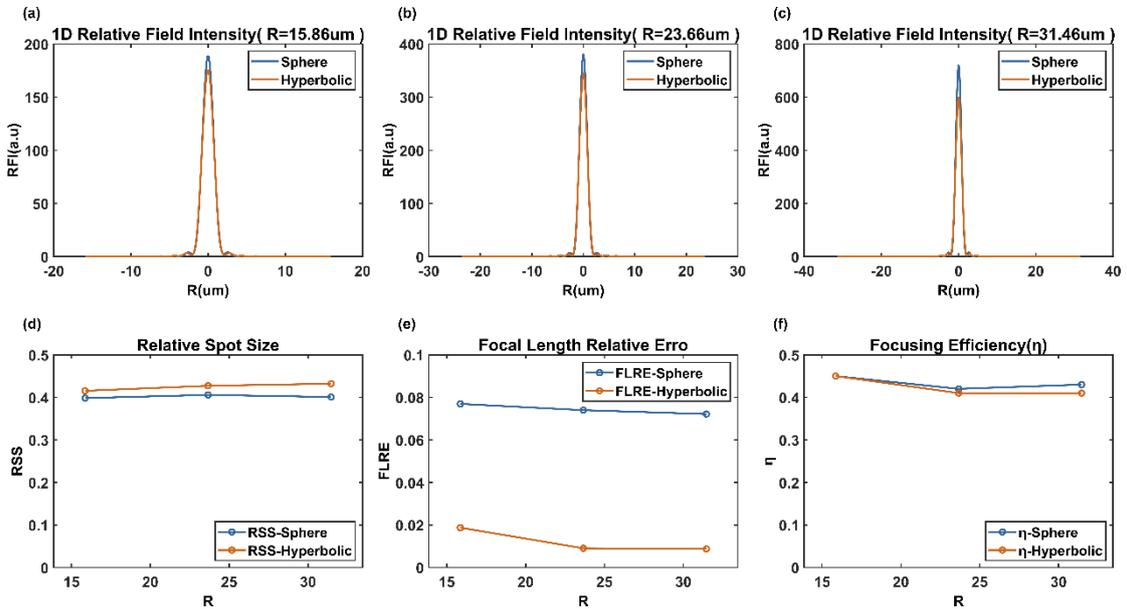

Fig. 12 Schematic of focusing effects for spherical and hyperbolic phase profiles with varying radii *R* (fixed numerical aperture). (a)-(c) show the 1D relative field intensity of focal points for both phase profiles at different lens radii. (d) shows the relative spot size (RSS) versus radius for both phases. (e) shows the focal length relative error (FLRE) versus radius for both phases. (f) shows the focusing efficiency (η) versus radius for both phases.

As shown in Fig. 12, as the lens radius increases, the spherical phase profile exhibits a smaller full width at half maximum (FWHM) and a higher peak relative field intensity compared to the hyperbolic phase profile. This demonstrates that the spherical phase profile possesses stronger light-convergence capability than the hyperbolic phase. In contrast, the full width at half maximum (FWHM) of the hyperbolic phase profile increases with lens radius, demonstrating that the hyperbolic phase profile exhibits spherical aberration. $R_1$=15.86 um,FWHM$_{hyperbolic}$ =1.763 um; $R_2$=23.66 um, FWHM$_{hyperbolic}$ =1.813 um; $R_3$=31.46 um, FWHM$_{hyperbolic}$=1.834 um. Calculation shows: $R^2$=0.95, σ =0.036 um. Proved the strong correlation between hyperbolic phase and radius.

In contrast, the full width at half maximum (FWHM) of the spherical phase profile remains approximately: $R_1$=15.86 um，$FWHM_{sphere}$=1.69 um; $R_2$=23.66 um，$FWHM_{sphere}$=1.722 um; $R_3$=31.46 um，$FWHM_{sphere}$=1.701 um. Calculation shows: $R^2$=0.098, σ =0.015 um. Proved the noncorrelation between spherical phase and radius.

For metalens with spherical and hyperbolic phase profiles at radii $R_1$=15.86 um，$R_2$=23.66 um and $R_3$=31.46 um the spherical phase demonstrates consistent performance advantages. While both designs achieve comparable focusing efficiencies at $R_1$ (0.45), the spherical phase exhibits marginally higher efficiencies at larger radii (0.42 vs. 0.41 for $R_2$; 0.43 vs. 0.41 for $R_3$). More strikingly, the spherical phase delivers significantly enhanced focal energy, outperforming the hyperbolic design by 7.6% (188.456 a.u. vs. 175.18 a.u..), 10.6% (380.497 a.u. vs. 344.193 a.u.), and 20.4% (719.8 a.u. vs.597.856 a.u.) for $R_1$, $R_2$, and $R_3$, respectively. This monotonic increase in the peak relative field intensity enhancement percentage with radius provides direct verification of our theoretical model. It demonstrates that the hyperbolic phase fails to effectively converge light at the lens periphery, resulting in a rising ratio of spherical-to-hyperbolic peak relative field intensity as the radius expands. Such reduced focusing capability of the hyperbolic phase at the lens edge constitutes definitive evidence of spherical aberration. This energy gain scales with lens size, suggesting that the spherical phase mitigates geometric aberrations more effectively in extended metalens. Combined with its reduced FWHM (4.1–7.3% smaller than the hyperbolic phase), the spherical phase achieves superior spatial resolution and energy concentration without compromising focusing efficiency, positioning it as a robust alternative to conventional hyperbolic designs.

$R_3$=31.46 um was selected for convergence analysis as it maximizes spherical aberration sensitivity in hyperbolic phase lenses (Fig. 12), providing a conservative assessment. Grid convergence was confirmed by coarsening the minimum mesh step from 0.00025 um to 0.01 um (40× relaxation).

Key observations:
- Spherical phase: ΔFWHM<0.06% (radius-insensitive focusing)
- Hyperbolic phase: $R^2$=0.95 preserved (consistent spherical aberration)
- All metrics varied <1.2% (Supplementary Section6, Table S1)

Theoretically, the spherical phase profile is free of spherical aberration, and the full width at half maximum (FWHM) of its focal spot remains constant regardless of radius. The observed deviation—where the FWHM of the spherical phase profile does not strictly remain constant—arises from mismatches between the metalens phase distribution and the meta-atom phase library, as well as the lack of optimization for the positions of spherical isophase surfaces.

## 4. Conclusion

This work establishes an isophase surface interpretation of metalens, deriving a spherical phase profile defined as the difference between outgoing spherical and incident planar isophase surfaces. We introduce a normal-tracing method to predict spherical aberration during phase design, validated by simulations comparing spherical and hyperbolic profiles. Three test groups demonstrate consistent advantages of the spherical phase: Fixed radius/variable numerical aperture *NA*; Fixed focal length/variable radius *R*; Fixed NA/variable radius *R*. In the third simulation (NA=0.317), the full width at half maximum of the spherical profile is 4.1-7.3% smaller than that of the hyperbolic design, and the peak relative field intensity (PRFI) is higher. At NA=0.317 with increasing radii ($R_1$=15.86μm, $R_2$=23.66μm, $R_3$=31.46μm), spherical phase maintains comparable efficiency while delivering significantly stronger focal energy: +7.6% ($R_1$), +10.6% ($R_2$), and +20.4% ($R_3$) enhancement. This radius-dependent PRFI gain directly verifies our model and confirms hyperbolic phase's inability to effectively converge peripheral light—definitive evidence of spherical aberration.

The spherical phase fundamentally decouples phase modulation from geometric constraints, (NA<0.707). Its superior aberration control provides higher resolution and energy concentration without efficiency compromise, establishing it as a robust alternative to conventional hyperbolic designs. Future work will target achromatic spherical phase metalens and optimization of the axial distance between spherical isophase surfaces and the metasurface.

Data Availability:
MATLAB code for aberration characterization via equiphase ray tracing is available at https://github.com/YXHujs2212103076/Spherical-Phase.git
Full FDTD parameters are documented in Supplementary Section 7, Table S2.